\documentstyle[12pt]{article}
\newcommand{\be}{\begin{equation}}
\newcommand{\ee}{\end{equation}}
\newcommand{\bea}{\begin{eqnarray}}
\newcommand{\eea}{\end{eqnarray}}
\newcommand{\nn}{\nonumber}

\begin{document}
\vspace{1.0cm}
\begin{center}
{\bf \large ZERO MODES OF  FIRST CLASS SECONDARY CONSTRAINTS \\
\vspace*{0.5cm}
                           IN GAUGE THEORIES}\\

\vspace{1.0cm}
{\sc A. Khvedelidze,  V. Pervushin}
\vspace{0.5cm}

Bogoliubov Theoretical Laboratory, Joint Institute for Nuclear Research,\\
Dubna, Russia

\end{center}
\vspace{1.0cm}
\begin{abstract}
{ Zero  modes  of first  class  secondary constraints in
  the two\---dimensional  electrodynamics  and  the four\---dimensional
   SU(2) Yang-Mills theory  are considered
   by the method of reduced phase space quantization in the context
  of the  problem of a stable vacuum. We compare the description of
   these modes in  the Dirac extended method  and reveal their
   connection with the topological structure of the gauge symmetry
   group.  Within the framework of the "reduced" quantization we
   construct a new  global realization of the homotopy group
   representation in the
   Yang\--Mills theory, where the role of the stable vacuum with
   a finite action plays the Prasad\--Sommerfield solution .}
\end{abstract}

\vspace{1.5cm}

\centerline{\bf Introduction}
\vspace{0.3cm}

There   is  a  significant   difference   between the  description    of
Yang\---Mills field ground state  and that of any   other fields  for
which the conventional  methods of quantum field theory  work.
The instability  of the naive perturbation theory (for example, see ref.
[1] and review [2]) is one of the crucial problems in
application of the non-Abelian field theory to  hadron physics.

 Efforts  to understand the   problem  of the QCD vacuum   give  rise   to
a huge  number of  speculations   connected  with    the  nontrivial
topological   properties     of  gauge fields [3\--8].

In this article we want to continue these attempts from  a point  of view
of the reduced phase space quantization [8], where only the physical gauge
field components are included in the canonical scheme. The 'reduced'
approach helps to reveal the significant role of zero
modes of the  secondary first\---class constraints ( by the definition
in refs.~\cite{Dirac,Slavnov}) in  the extended quantization
of gauge theories. We will demonstrate that the presence of
zero--modes  reflects  a  global  structure  of
the  initial   gauge symmetry group. These zero modes can be also
treated as some collective excitations of gauge fields.
An example of  that type zero mode is the Coleman electric field in  the
two\---dimensional QED \cite{Coleman,Ilieva,Jackiw}.
It is well  known that the local $U(1)$  in the two\---dimensional  space
time and the non\---Abelian compact  groups in  four dimensions  have the
same topological properties.    Therefore,    before   the
consideration  of  the  non\---Abelian  gauge  theory  we  study  a
simple example,  electrodynamics on a finite line  and emphasize
the  nature  of  these  zero  modes  as  remaining  quantum mechanical
variables.  With  this experience we  proceed to  investigate the
four\--dimensional $SU(2)$ gauge model  where in direct analogy  with the
previous example  we introduce the same type residual variable  describing
the zero mode of the secondary  constraint.  On the basis of  this , we
speculate on a possible role  of these collective excitations for  the
Yang\---Mills ground state and stable perturbation theory.

The paper will be organized as  follows.
In section 1 , we present  a systematic analysis of electrodynamics  in
two\---dimensional finite  space time  in the Lagrangian  and Hamiltonian
forms.   Section  2  is  devoted  to  the  (1+3)  dimensional $SU(2)$
Yang--Mills  theory.   We  prove  a  no--go  theorem  about the local
 realization  of  the  representation  of  a homotopy group without the
collective mode,  and show  that the  presence of a zero\--mode of  the
first--class  secondary  constraint  leads  to  another   realization
different from the "instanton" one  [3-5].

\vspace{1.5cm}

\section{Electrodynamics ~ in~ the~ two -- dimensional~ finite
space --  time}

\subsection{Zero mode of the Gauss equation }

Let us start with the Abelian \( U(1) \) gauge theory action
in  the two-dimensional finite space time
\begin{eqnarray}
W\big[A^\mu \big]  = \int \limits_{-\frac{1}{2}T}^{\frac{1}{2}T}dt
\int \limits_{-\frac{1}{2}{R}}^{\frac{1}{2}{R}}
dx \left( -\frac{1}{4} {F^2}_{\mu\nu} \right). \label {eq:action}
\end{eqnarray}

\noindent In the (1+1) space time we have only electric tension $E$~~
\footnote{\normalsize{Below in the text we will use the following notation
$\dot{f}:\;=\frac{\partial {f}}{\partial t },
\partial{f}:\;=\frac{\partial{f}}{\partial x}.$}}
\be
F_{01}:= \dot{A_1}-{\partial{A_0}}\; =\;E .
\ee
\noindent The action  is invariant under the local gauge transformation
\be
A_\mu (t,x) \,\,\to\,\,  A^\prime _\mu(t,x) = g(t,x)
( A_\mu (t,x)+ \frac{i \hbar}{e} \partial _\mu) g^{-1}(t,x), \label
{eq:gauge.tr}
\ee
affected by an element of the gauge group
\begin{equation}
g(t,x_1)=\exp\; (\frac {i}{\hbar}\; \lambda(t,x))
\end{equation}
with an arbitrary function $\lambda (t,x)$ .
(The constant $e/\hbar$ has the dimension of mass and
 $\lambda / \hbar$ is dimensionless.)

The Euler \--- Lagrange  equations for the
gauge field follows from the action (\ref {eq:action}) by varying
$A_{\mu}(t,x)$
\bea
{\partial}^2 A_0(t,x) - \partial \dot{A_1}(t,x)\;&=&\;0, \label{eq:Gauss} \\
\ddot{ A_1}(t,x) - \partial \dot{A_0}(t,x)\;& = &\;0 .      \label{eq:long}
\eea
The Gauss equation (\ref {eq:Gauss}) does not contain a time derivative of
the time component \( A_0 \) and is considered as a constraint.

The general  solution of  the Gauss law (\ref{eq:Gauss}) with respect
to the time component can be represented as  a
sum of a general solution of the homogeneous equation

\[ {\partial}^2 {\varphi}_0(t,x)\;=\;0 \;,\;\;
\varphi_0(t,x)\;=\;c_1(t)\;+\;c_2(t)\;x , \]
and a particular solution of the inhomogeneous one constructed with
the help of Green's function $G(x, x')$:
\be
A_0 (t,x)= {\varphi}_0(t,x) + \int\limits_{-\frac{1}{2}{R}}^{\frac{1}{2}{R}}
d{x}'G(x\;, x') \left ( \partial \dot{A_1}(t,{x}')\right)
\;,\label{eq:solution}
\ee
\begin{equation}
{\partial}^2 G(x\;, x')\;=\;\delta (x-x') . \label{eq: gf}
\end{equation}

To specify the zero mode \( \varphi_0(t,x) \) and the Green function
\(G(x\;, x'), \) we need a boundary condition for the gauge
field \(A_\mu(t, x) \). The usually exploited boundary condition
\be
A_\mu(t,x) \bigg\vert_{x=\pm \frac{1}{2}R}\;=\;0 \label{eq:zbound}
\ee
leads to the well\---known result for the zero mode \[ {\varphi}_0(t,x)\;=\;0
\]
and the Green  function

\[ G(x,\; x')\;=\;\frac{1}{2}\mid x - x' \mid \;+\;{(xx')
\over R}\;-\;{R \over 4}. \]
Substitution of the solution (\ref{eq:solution}) with these quantities into
(\ref{eq:long})  leads to the identity
\begin{equation}
\ddot{ A_1}(t.x) \;=\;\ddot{ A_1}(t,x) . \label{eq:ident}
\end{equation}
Thus , as have been expected , we do not get any restriction on $A_1(t.x)$ .
Due to the gauge invariance (\ref{eq:gauge.tr}) only
transversal components are dynamical. In the two\--dimensional space
we have only a longitudinal component. As a consequence , we obtain
\[E\;=\;0.\]

Now let us suppose a more subtle than (\ref{eq:zbound}) condition
\be
A_\mu(t,x) \bigg\vert_{x=+ \frac{1}{2}R}\;=
\;A_\mu(t,x) \bigg\vert_{x=-\frac{1}{2}R}\;+\;a_\mu(t) \label{eq:bound}
\ee
with an arbitrary time\--- dependent vector $a_\mu (t)$ and
\bea
{\dot{A}}_1(t,x) \bigg\vert_{x=+ \frac{1}{2}R}\;&= &\;
{\dot{A}}_1(t,x) \bigg\vert_{x=-\frac{1}{2}R} \\
{\partial{A}}_0(t,x) \bigg\vert_{x=+ \frac{1}{2}R}\;&= &\;
{\partial{A}}_0(t,x) \bigg\vert_{x=-\frac{1}{2}R}.
\eea
These conditions mean for the physical quantity
\(E\) the following:

\[
E \;\bigg\vert_{x=+\frac{1}{2}R} \;=\;E \;\bigg\vert_{x=- \frac{1}{2}R}.
\]

It is evident that for this case we have zero mode
\[ {\varphi}_0(t,x)\;=\;c_1(t)\;+\;{a_0(t) \over R}\;x .\]

Due to the presence of this zero mode of the operator \( {\partial}^2 \),
we face the problem of the correct  definition of  Green's function with
the boundary conditions
\bea
G(x,x') \bigg\vert_{x=+ \frac{1}{2}R}\;&= &\;
G(x,x') \bigg\vert_{x=-\frac{1}{2}R} \\
{\partial G(x,x')} \bigg\vert_{x=+ \frac{1}{2}R}\;
&= &\;{\partial G(x,x')} \bigg\vert_{x=-\frac{1}{2}R}.
\eea

To solve this problem , we exclude the zero mode from the Green function
spectral representation
\begin{equation}
G(x\;,x')\;=\;\sum_{n=-\infty, n\neq 0}^{\infty}\frac{1}{\lambda_n}
u_{n}(x)u^{*}_n(x'),
\end{equation}
where the function $u_n(x)$ is an eigenfunction of the one\--dimensional
Laplace operator with boundary conditions of the type (14), (15)

\[
{\partial}^2 {u_n(x)}\;=\;\lambda_n u_n(x), \;\;\;\;\;\;\;\;
 u_n(x)\;=\;\frac{1}{\sqrt R} \exp(i\frac{2\pi n }{R}x),
\]
with an eigenvalue \(\lambda_n \;= - (\frac{2\pi n }{R})^2 \).

The representation (16) leads to the following equation

\[
{\partial}^2 G(x, x')\;= \sum_{n=-\infty, n\neq 0}^{\infty}
u_{n}(x)u^{*}_n(x')
\;=\;\delta (x -x')- \frac{1}{2R}
\]
instead of  the conventional one (8). Nevertheless , it is
easy to check  the representation (\ref{eq:solution}) with this
solution of the Gauss equation (\ref{eq:Gauss}) because $A_1$ satisfies
(12).
The explicit form of the Green function (16) is

\[
G(x - x')\;=\;\frac{1}{2}\mid x - x' \mid - {(x - x')^2 \over 2 R}
- { R \over 12 }.
\]

After  substitution of the solution (\ref{eq:solution})
into the equation for $A_1(x)$ instead of identity (\ref{eq:ident}) we get
\begin{equation}
\partial \dot{\varphi_0}(t,x)\;-\;\frac{1}{R}
\int\limits_{-\frac{1}{2}{R}}^{\frac{1}{2}{R}} d{x}'
\ddot{A}_1(t,x')\;=\;0 .
\end{equation}

This is the crucial point. We obtain the  remaining
variable depending on the zero mode and on  the functional \( N_L[A_1] \)
\begin{equation}
N_L[A_1]\;=\;\frac{e}{2\pi\hbar} \int\limits_{-\frac{1}{2}{R}}^
{\frac{1}{2}{R}} d{x}'{A}_1(t,x') .
\end{equation}
It is useful to introduce the following notation for remaining variable

\begin{equation}
N_{T}(t)\;=\;N_0(t)\;+\;N_L[A_1],
\end{equation}
where
\begin{equation}
N_0(t) \;=\;\frac {eR}{2 \pi \hbar}\int \limits_{-\frac{1}{2}T}^{t}dt'
\partial {\varphi_0}(t',x) .
\end{equation}
In  terms of \(N_T(t) \) eq.(17) has a simple form
\begin{equation}
{\ddot {N}}_T(t) \;=\;0
\end{equation}
and for electric tension we get
\[E\;=\; \frac{2\pi \hbar}{R} {\dot {N}}_T(t). \]

It is easy to see that the new variable \(N_T(t) \) is connected with
the  two dimensional topological invariant ,
the Pontryagin number functional $\nu[A]$

\[
\nu[A] \;=\; \frac {e}{4 \pi \hbar} \int d^2 x \epsilon_{\mu\nu}F^{\mu\nu}\;=\;
\frac {e}{2 \pi \hbar}\int \limits_{-\frac{1}{2}T}^{\frac{1}{2}T}dt
\int \limits_{-\frac{1}{2}{R}}^{\frac{1}{2}{R}}dx
(\dot A_1 -
\partial  A_0)
\]
in a simple way
\be
\nu[A] \bigg\vert_{\mbox{constraint}} \,=\, N_T(\;\frac{T}{2}\;)
- N_T(-\frac{T}{2}\;).
\ee

Now, we can see the double role of the remaining variable  $N_{T}(t)$ in
different gauges.
In the gauge $A_1\;=\;0$ it looks like a zero mode of the Gauss equation for
electric tension
\[
{\partial}^2  A_0(t,x) \;=\;0 .
\]
In the temporal gauge  $A_0\;=\;0 \;~~, N_{T}\;=\;N_L[A_1]$ is the
topological variable which transforms under the
residual  stationary gauge transformations
\[
A_1 (t,x) \,\,\to\,\,  A^\prime _1(t,x) = g(x) ( A_1(x_1)+
           \frac{i \hbar}{e} \partial ) g^{-1}(x), \;\;
\;\;g(x)=\exp\; (\frac {i}{\hbar} \lambda(x)\; )  \label {eq:gauge.str}
\]
in the following way:
\be
N_L[A^\prime _1]\;=\;N_L[A_1]\;+\;
\frac{\hbar}{2\pi}\left[ \lambda(\frac{R}{2})\;
-\lambda(-\frac{R}{2})\right] .
\ee

Recall that the functions $g(x) $ represent
maps of the line $\left (-R/2,\;R/2\right)$ with the
identified ends $g(\frac{R}{2}) \;=\;g(-\frac{R}{2})$ onto the $U(1)$--group
space. All these maps are split into the
classes characterized by the integer index $n$ pointing out how
many times the closed line turns around the
$U(1)$--space circle.
The quantity
$g_{(n \neq 0)} $ is called the large gauge transformation; while
$g_{(n = 0)} $ , the small one.

The factor--space of all stationary gauge transformations with
respect to the  small ones $G/G_0$ coincides with the homotopy group of all
one--dimensional closed paths on the $U(1)$--circle
\begin{eqnarray}
\Pi _1 \left ( U(1)\right ) = Z .
\end{eqnarray}
where $Z$ is the group of integers.
The new variable \(N_T(t) \) is invariant under the small gauge
transformations and changes by an integer under large transformations.
\begin{equation}
N_T(t) \to N_T + n  .
\end{equation}

The invariance of the theory  under the large gauge transformation
means that the points $N_T$, $N_T + n$ are physically identical.
The configurations $N_T\;=\;0$ and $N_T\;=\;1$ are the same; so the manifold
\( \{ N_T \} \) is a circle of the length of unity.

Thus, the  explicit solution of the constrained equation (\ref {eq:Gauss})
leads to  some reduced theory, effective action for which can be obtained
after substitution of the solution (\ref{eq:solution}) into initial action (1)
\begin{eqnarray}
W\big[A_\mu\big]\bigg\vert_{\mbox{constraint}} = W^{Red}\big[N_{T}(t) \big] & =
&\int \limits_{-\frac{T}{2}}^{\frac{T}{2}}dt
\left( \frac{1}{2}{\dot{N}_{T}(t)}^2 I \right) .
\end{eqnarray}
The gauge field theory in the 1+1 space time reduces to a simple mechanical
system .
The  reduced action with the definition of the manifold \( \{N_T \} \)
describes a   plane rotator with the mass
\(I\;=\;\frac{2\pi}{eR}\).

To get a physical consequence of the new  variable , we should quantize
our reduced theory.
Thus , the next goal  is the  consideration of this zero mode in
the Hamiltonian form of gauge dynamics.

\vspace{1.cm}

\subsection{  Zero modes in the Hamiltonian approach.
              Primary and secondary reduction }

It is  easy to get the Hamiltonian form of  the reduced
theory starting  from the reduced action (26)
\begin{eqnarray}
W^{Red} = \int^{T/2} _{-T/2} dt \left[ P \dot N_T - \frac {1}{2}\;
P^2 I^{-1} \right] ,
\end{eqnarray}
where $P$ is canonically conjugate to $N_T$.
For the generalization to the non-Abelian case , it is useful
to obtain this action and to elucidate the status of zero modes
in the canonical Hamiltonian scheme.

For our purpose it is convenient to use first order formalism action
\be
W_{I} \big[E, A^\mu\big] =
\int_{0}^{T}dt \int_{-\frac{1}{2}{R}}^{\frac{1}{2}{R}}dx
\left(E(\dot A_1 - \partial A_0) - \frac{1}{2}E^2 \right) ,
\label {eq:action1}
\ee
where the time component $A_0$ plays the role of the Lagrange factor.
The momentum canonically conjugate to $A_0(t,x)$  is equal to zero
\be
{\cal {\pi}}_0(t,x) \;=\;\frac{\partial {\cal L}}
{\partial \dot{A^0}(t,x)}\;=\;0 .
\ee
Equation (28) is the  primary constraint
and  its Poisson bracket with the canonical Hamiltonian \({\cal H}\) leads to
a secondary constraint for the electric tension
\begin{equation}
\{ {\cal \pi}_0,\; {\cal H} \} \;=\; {\partial} E \;=\;0  .
\end{equation}
These two  constraints according to Dirac's definition [9] , form the
first-class ones.
Equations (29) and  (30) mean that the time component is not physical and
can be removed from phase space by the  gauge transformation.
The removal of nonphysical components by explicit solving of primary
and secondary constraints will be called the {\it primary and
secondary reductions\/} respectively.
To remove the time component , it is enough to choose the gauge $A_0=0$ .
The primary reduction of the  action (28) gives

\begin{eqnarray}
W^{Red}_I = \int d^2 x \left[ E \dot A_1 - \frac{1}{2} E^2 \right]
\end{eqnarray}
with the  Gauss law \[ \partial E \;=\;0 . \]

This first-class secondary constraint should be accompanied by the
second \--class constraint (gauge condition). Let us choose
the gauge
\be
\partial \dot A_1\;=\;0 .
\ee
This gauge is distinguished by the equation of motion following from the
action (31)
\be
E\;=\; \dot A_1
\ee
and by the constraint (30) .

The secondary reduction of the Hamiltonian scheme action (28)
evidently coincides with
the action (26) obtained from the Lagrangian approach.
To verify this , let us  write down the
explicit solution of constraints (30) and (32) in the form
\begin{eqnarray}
E &=& P \frac {e}{2\pi \hbar}\;+\;E^T,\;\;\\
A_1& =&  g(N_0) ( A_1^T + \frac{i \hbar}{e} \partial) g^{-1}(N_0) ,
\end{eqnarray}
where the gauge transformation
\begin{eqnarray}
g(N_0) = \exp \left\{ i \, \frac{2\pi x}{R}\; N_0 (t) \right\}
\end{eqnarray}
and  $ A^T $ and $ E^T $ are the transversal variables equal to zero in the
two\--dimensional case.

Thus, a global subgroup of gauge symmetry leads to collective
excitation $P$ of the type of a zero  mode of the first class secondary
 constraint (30) which is  the remaining
longitudinal part of the gauge field momentum. This zero--mode
is accompanied by the zero mode of the "radiation" gauge, which is well known
as
the Gribov ambiguity \cite{Gribov}.

\vspace{1.5cm}

\subsection {Quantization of the reduced theory}
\vspace{0.3cm}

For quantum description of the reduced theory we will use
a fixed time  Schr\"odinger representation.The
canonical variables \({\hat{N}}_T(t), {\hat{P}}(t)\) are fixed at time
\(t^*\)
\[ {\hat{N}}_T\;=\;{\hat{N}}_T(t^*),\;\;\;\;\; \hat{P}\;=\; {\hat{P}}(t^*) \]
and satisfy the commutation relation
\begin{equation}
i \left [ \hat{P},{\hat{N}}_T \right ] = \hbar .
\end{equation}

The  stationary  Schr\"odinger equation
\begin{eqnarray}
H \Psi _{\epsilon} = \epsilon \Psi _{\epsilon},
\end{eqnarray}
should be complected by the constraint of identification
of the points $N,\;\; N + 1$ on the  circle
\begin{eqnarray}
\Psi_{\epsilon}\,(N + 1) = e^{i\theta}\,\Psi_{\epsilon}\,(N) \;\;\;\;\;\;\;
(0 \leq \theta \leq \frac {\pi}{2})  .
\end{eqnarray}
The solution of these equations is the Bloch plane wave
\begin{eqnarray}
\Psi_{\epsilon} = e^{\frac{i}{\hbar}P N }, \;\;\;\;\;
P = (2\pi k + \theta) \hbar ,
\end{eqnarray}
where $k$ is the number of the Brillouin zone. The spectra of the
electric tension  and the energy
have the following forms
\begin{eqnarray}
E = e\,\left( k + \frac{\theta}{2\pi}\right ) \\
\epsilon = RE^2\,/\,2
\end{eqnarray}
which coincide for the ground state  ($k\;=\;0$) with the Coleman constant
electric tension ~\cite{Coleman}. The nonzero tension (or the collective
persistent current in the functional space) appears here as a pure
quantum effect (of the type of the Josephson one) due to the
jump of the phase of the wave function (40).

There is another way of describing this zero  mode
where it is represented as the functional of the local
variable [4,5]. However, different (equivalent for the
considered model) ways of  introducing  the
zero  mode lead to different results in the non\--Abelian
theory. The reader can see this fact in the next section.

\vspace{1.5cm}

\section{ The $SU(2)$ Yang\---Mills Theory in Four Dimensions }

\vspace{0.3cm}

\subsection{ Primary reduction\/}

We start with primary  reduction of the Yang \--- Mills theory with the
local  $SU(2)$ group
\begin{equation}
W^{Red}\; =\; \int d^4x \left[ E^a_i{\dot{A}}^a_i-
\frac{1}{2} \left( {E^{a}_{i}}^2 + {B^{a}_{i}}^2
\right)\right],
\end{equation}
where the electric tension \( E^a_i \) satisfies the secondary constraints
\begin{equation}
       \nabla ^{ab}_i E^b_i\;=\; \left(\delta^{ab}\partial_i\ +
e \epsilon^{acb} \,A^c_i\right) E^b_i\;=\;0,
\end{equation}
and  magnetic tension \(B^a_i\)
\begin{equation}
B^a_i = \epsilon_{ijk}\; \left( \partial_j A^a_k + \frac{e}{2}\,
\epsilon^{abc}\,A^b_j\,A^c_k \right) .
\end{equation}
Below we will use the gauge
\begin{eqnarray}
\nabla^{ab}\,_i\, (A) \dot A^b_i = 0
\end{eqnarray}
which is consistent with the equations of motion
\begin{equation}
E^a_i\;=\;{\dot{A}}^a_i .
\end{equation}

This theory has the  topological nontrivial gauge symmetry group
{}~[3-8]
\begin{eqnarray}
\hat A_\mu\,\,\to\,\, \hat A_\mu^g = g (\hat A_\mu +
\partial_\mu)\,\,g^{-1};\;\;
\hat A_\mu = \frac{e\tau^a A^a_{\mu}}{2i} .
\end{eqnarray}
All stationary transformations with the boundary conditions
\begin{eqnarray}
\lim \;g\;\;(\vec x) = 1\;\;\;\;
\mid {\vec  x} \mid~~~ \to~~~ \infty
\end{eqnarray}
represent the manifold of three--dimensional closed paths on the
three--dimensional sphere SU(2), and can be   split into the
classes characterized by the integer index of a map (n) of the space
$\{\vec x\} $ into the SU(2) group space:
\begin{eqnarray}
n = \frac{1}{24\pi^2}\;\;\int d^3 x\; \epsilon^{ijk}
tr \left[ \hat V _i\; \hat V _j\; \hat V _k \right];\;\;
\hat V_i = g\; \partial\,_i g^{-1} .
\end{eqnarray}
As in eq.(24), we can speak here about the homotopy group
\begin{eqnarray}
\Pi_3 (SU(2)) = Z .
\end{eqnarray}
There  is  a topological variable $N_L[A]$
\begin{eqnarray}
N_L[A] = \frac{e^2}{16\pi^2}\;\; \int d^3 x \epsilon_{ijk}
\;(A^a_i\; \partial_j\; A^a_k + \frac{1}{3}\; \epsilon^{a b c}
\, A^a_i\, A^b_j\, A^c_k)
\end{eqnarray}
which realizes the representation of the homotopy group ~ \cite{Faddeev}
\begin{equation}
N_L[A^g]\;=\;N_L[A]\;+\;n .
\end{equation}

\vspace{1.5cm}

\subsection{ No\--Go Theorem for the Local Quantum Representation
of the Homotopy Group}
\vspace{0.3cm}

Apart from the experience obtained from the above considered two-dimensional
theory there is one more mathematical argument in favor of the
existence of the independent collective variable $N_T$ of the type
of (19).

The exact formulation of the problem of quantization of the
Yang--Mills theory with the nontrivial homotopy group
is given in refs.~\cite{JackiwRebi,Callan} and consists in solving
of the set of equations
\begin{eqnarray}
H_L\Psi_{\epsilon}& = &\epsilon \Psi _{\epsilon}\\
\nabla_i E_i\;{\Psi}_\epsilon & = &0 \\
T_L\Psi_{\epsilon}& = &e^{i\theta}\,\,\Psi_{\epsilon} .
\end{eqnarray}
A first equation is the stationary  Schr\"odinger equation with the
Hamiltonian
\begin{equation}
H_L[A,E] = \int d^3 x \frac{1}{2}\,(E^{a2}_i + B^{a2}_i) ,
\end{equation}
Eq.(55) reflects the invariance of the theory under the small gauge
transformations, and Eq.(56) describes the covariant properties of the
wave function  under  a large gauge transformation.
The topological shift operator $T_L$ has the form
\begin{equation}
T_L = \exp\left\{ { d \over {dN_L[A]}} \right\} ,
\end{equation}
where $N_L $ is the functional (51).
This form is justified in refs.\cite{JackiwRebi,Callan} by
representing  the solution of (54)--(56) in the form
of the Bloch wave function
\begin{eqnarray}
\Psi_{\epsilon} (N_L,\,A^T) = e^{\frac{i}{\hbar} P\;N_L}\;
\Psi_{\epsilon} (A^T) \nonumber
\end{eqnarray}
and by the exact nonphysical solution with  energy \(\epsilon\;=\;0 \)

\begin{eqnarray}
\Psi_o = \exp\left\{ \pm \frac{8\pi^2}{e^2}\,N_L [A] \right\} . \nonumber
\end{eqnarray}

\vspace{0.5cm}

\noindent \underline{No\--go theorem}:
{\bf There are no physical solutions of equations}
(54)--(56) .

The Proof: It is easy to check that the operators
$H_L, T_L $ do not commute
\begin{eqnarray}
[H_L,T_L] \ne 0;\;\;\; \left[ [H_L [H_L,\;T] \right] \ne 0;\;\;\; \nonumber
\end{eqnarray}
therefore they cannot have a complete system of physical eigenstates.

In the two\-- dimensional Abelian case , this local realization
works due to the absence of a transversal variable.
There is only a  plane wave excitation.
In the three\-- dimensional case , these transversal variables
describe the oscillator\-- like local excitations in the Schr\"odinger
equation  due to the  magnetic
field potential  while (56) means that the wave function is
simultaneously a plane wave, which is impossible.

\vspace{1.5cm}
\subsection{Secondary Reduction}
\vspace{0.3cm}

As we have seen in two-dimensional case, there is another way to get a
nontrivial physical representation of the homotopy group (51).
For this goal it is  sufficient to introduce an independent collective
topological variable  $N_0$,  which describes the Gribov ambiguity
of the " motion equation gauge" (46)
\be
\hat A_i = {\hat A_i}^{g_{N_0}} = g_{N_0}\,\left( \hat A^T_i +
\partial_i\right )\;{g_{N_0}}^{-1},
\ee
and its conjugate momentum
\be
\hat E_i = g_{N_0} \left[ \hat E^T_i + P_0 I_B {I_{\Phi}}^{-1}
{\nabla}_i \; (A^T)  \hat {\Phi}_0 \right] {g_{N_0}}^{-1}
\ee
as a \underline{zero mode of the first-class secondary constraint}
(44).
Here ${\Phi}_0$ is the zero eigenfunction of the Gauss constraint (44)
\be
{\nabla}^{ab}_i (A^T)\; {\nabla}^{bc}_i\, (A^T)\; {\Phi_0}^c = 0
\ee
\be
g_{N_0}\;=\; T \exp \left( \hat {\Phi}_0 N_0(t){I_B}^{-1}\right)
\ee
$I_B$ , $ I_\Phi$ are the following surface integrals:
\begin{eqnarray}
I_B &=&\int d^3 x (\nabla_i \Phi)^a \bar B^a_i
\equiv \int d^3 x \partial_i (\bar B^a \Phi^a);\;\;
\bar B = \frac{e^2}{8\pi^2}\, B^a_i  \nonumber \\
I_\Phi& = &\int d^3 x (\nabla_i \Phi)^a
(\nabla_i \Phi)^a = \frac{1}{2} \int d^3 x \partial^2_i (\Phi^a)^2 .
\end{eqnarray}

Note that the new variables $E^T$ and $ A^T$ satisfy the same
constraints(44),(46) while the topological variable (50) and action
(43) acquire additional terms
\be
N_T[A^T, N_0]\;=\;N_L[{A^T}^{g_{N_0}}]=N_L[A^{T}]\;+ N_0 \;+{\mbox{Inv.term}}
\ee
\begin{eqnarray}
W^{Red} \big[ A, E \big]  =  W^{Red} \big[ A^T, E^T  \big ] \;+\;
\int \limits_{\frac{-T}{2}}^{\frac{T}{2}} dt \left[ P \dot N - \frac {1}{2}\;
P^2 I_\Phi {I_B}^{-2} \right]
\end{eqnarray}
Eq.(64) is defined  within a  term invariant under  large gauge
transformations; Eq.(65) is just the secondary reduction action.
Let us  consider the simplest case  when the surface integrals (63)
are time independent. We  choose  them as
\be
I_\phi =\frac{2 (2\pi)^3 }{\mu e^2}, \;\;\;\;\;  {I_B}\;=\;1
\ee
with $\mu $ being the parameter of the mass dimension.
Emphasize that this condition means a slow increase in the fields at
spatial infinity. An example of  fields like those is the well known Prasad-
 Somerfield solution ~\cite{22} of the Bogomolny equation
\begin{equation}
\nabla^{ac}_i (A_{asympt}){\Phi_0}^c = \frac{2\pi}{\mu}B^a_i (A_{asympt}),
\end{equation}
where
\begin{eqnarray}
{A_{asympt}}^a_i &=& \frac{1}{e} \epsilon^{abi} m^e
\left[\frac{\mu}{\sinh (\mu r)} - \frac{1}{r} \right];
\;\;\; m^l = \frac{x^l}{r}; \;\; r = \mid \vec x \mid \nonumber \\
(\Phi^a)_0& =& \frac{2\pi}{e} m^a \left [ \mu \coth{(\mu r)}
- \frac{1}{r} \right] .
\end{eqnarray}
For these fields , the invariant term in Eq.(64) has the following form:
\begin{equation}
\mbox{Inv.term}=-\frac{\sin (2\pi N_0)}{2\pi} .
\end{equation}
Thus , we get the non-Abelian analog of the Coleman
electric field in the (1+1) QED.

\subsection{The global representation of the homotopy group}

 From the reduced action (65) we get the following Hamiltonian:
\be
H_{Red}[P,E,A^T,E^T] \;=\; \frac{1}{2I_\Phi} {\hat P}^2 + H_L[A^T,E^T] ,
\ee
where $H_L$ is defined by (57).
In this case , the Schr\"odinger equation
\begin{eqnarray}
H_{Red} \; \Psi_{\epsilon} = \epsilon \Psi_{\epsilon}
\end{eqnarray}
admits the factorization of the wave function  on the plane wave
describing the topological collective motion, and oscillator like part
depending on transversal variables.
\begin{eqnarray}
\Psi_{\epsilon} (N_0,\,A^T) = e^{\frac{i}{\hbar} P N_0}\;
\Psi_{L} [A^T] .
\end{eqnarray}
Thus , the representation of homotopy group is realized as
\begin{eqnarray}
T_G \Psi_{\epsilon} = e^{i\theta} \Psi_{\epsilon};\;\;
T_G =\exp \left(\frac{i}{\hbar} \hat P \right)\; =\; \exp\; (\frac{d}{dN_0}).
\end{eqnarray}

\noindent Recall that  $P$ has a discrete spectrum
\begin{eqnarray}
P = (2\pi k + \theta)\; \hbar . \nn
\end{eqnarray}
The oscillator like part of the wave function  $\Psi_{L} [A^T] $ is
described by the Hamiltonian $H_L$.

It is useful to separate the stationary asymptotic part of the transversal
variable $A_{asympt}$
and the quasiparticle excitations with the zero boundary conditions
\begin{eqnarray}
\hat A^T (x_0, \vec x) = \hat A_{asympt}(\vec x) + \hat a^T (x_0; \vec x) .
\end{eqnarray}

In the "homogeneous" approximation, if we neglect
quasiparticles $\hat a^T (x_0; \vec x)$, we get from (65)
the following effective action
\begin{eqnarray}
W_{\rm {eff}} = W^{Red}[A_{asympt},E^T=0]\;=\;
\int \limits_{\frac{-T}{2}}^{\frac{T}{2}} dt \left[  \frac {1}{2}\;
P^2 I_\Phi {I_B}^{-2} -\int d^3x {\hat B}^2 \right] .
\end{eqnarray}
For the Prasad\--Somerfield asymptotic field there are  values of
the coupling constant
\begin{eqnarray}
\frac{e^2}{4\pi} = 1 / (k+ \frac{\theta}{2\pi})
\end{eqnarray}
for which the effective collective action (75) is equal to
zero.

We want to emphasize the attractive peculiarities of the considered global
realization, zero action and stability of perturbation theory under
small deformations \cite{Akhouri,Leutwyler}.
This  is just the main difference from instanton contributions witch are
suppressed by the action factors.
\vspace{1.5cm}

{\bf Conclusion}
\vspace{0.3cm}

We have discussed the mathematical and physical arguments in the favour of
the introduction of the zero modes of  the first\--class secondary
constraints in gauge theories.

It is shown that the reduced action approach  allows us to take explicitly
into account  the zero modes including the Gribov ambiguity mode
and clarifies their  double role as independent variable or the winding
number  functional.
To reproduce this result in the  Dirac  Hamiltonian approach ,
it is useful to consider the procedures of  primary and secondary
reductions. The primary reduction has been introduced by Dirac
to conserve the uncertainty principle for gauge field components
 included in the phase space (see discussion of eq.(2.28) in Dirac's
Lectures [9]). This reduction is equivalent  to the choice of a temporal
gauge. The secondary reduction consists in the fixation of the remaining
ambiguity due the presence of  stationary gauge transformations, generated
by the Gauss constraint. For this purpose , we explicitly solve the Gauss
constraint and  the additional gauge condition  that does not contradict
the equation of motion. At this step , to reproduce the result of
the Lagrangian method , it is necessary to introduce the {\it
zero mode of the first\--class secondary constraint\/} together with the
zero mode of the "motion equation  gauge " (second\-- class constraint).
These two modes are considered as canonically conjugate variables, the
winding number and its momentum.
The introduction of  independent modes allows us to consistently describe
the representation of the homotopy group and to construct the quantum
theory with  the effective finite action for the local excitation.

In the Yang\--Mills theory the zero\-- mode dynamics is realized in the
presence of  a stationary condensate of the type of the
Prasad\--Sommerfield "bag" [7,15,18] . Perturbation theory around this
condensate is stable unlike the conventional one.
This situation is very similar to the theory of gravity ,where the
metric excitation of the type of the Friedmann expansion leads
to the stabilization of the Universe. This metric
excitation is also the zero mode  sector of secondary constraint
in the theory of gravity [19].\vspace{0.3cm}

{\bf Acknowledgment }

\vspace{0.3cm}

The authors thank Profs. S.Sawada, F.Steiner, M.Volkov and Dr. H.Grosche
for useful discussions.One of the authors (V.P.) thanks the Russian Found
of Fundamental Investigations,Grant N 93\--02\--14411 for the support.

\end{document}